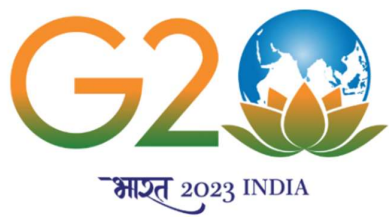
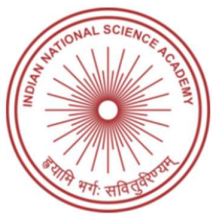
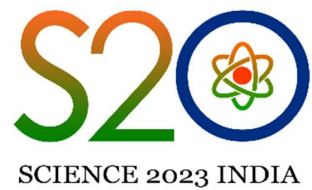

# Policy Brief

## WORKFORCE DEVELOPMENT IN ASTRONOMY AND ASTROINFORMATICS

Discussed and drafted during S20 Policy Webinar on Astroinformatics for Sustainable Development held on 6-7 July 2023

Contributors: Kartik Sheth, Kevin Govender, Vanessa McBride, Laura Trouille, Puji Irawati, Rana Adhikari, Rafael Santos, Paula Coehlo, Guiseppe Longo, Pranav Sharma, Ashish Mahabal

INDIAN NATIONAL SCIENCE ACADEMY – CENTRE FOR SCIENCE POLICY AND RESEARCH

**Introduction**
The discipline of astronomy and astroinformatics is dynamically evolving thereby creating a compelling opportunity to foster a more inclusive, diverse, and proficient workforce. This is crucial for addressing multifaceted challenges that emerge as we progress and harness the potential therein. To realize this goal, it's imperative to cultivate strategies that promote inclusive practices in STEM education, encourage participation from historically excluded groups, provide training and mentorship, as well as provide active champions, especially for students and early career professionals from (historically) excluded groups.

One must actively and intentionally work towards increasing the representation of successful scientists and STEM professionals that reflect the demographics of a nation at all career levels - such representation can have a transformative impact when early career scientists, especially those from (historically) excluded groups, see others like themselves successful in the field.

Currently the demographics in the field of astronomy and astroinformatics do not reflect the population in most nations. For example, in the United States, from 2018-2020, men earned 64% of the PhDs[1]. This fraction is not dissimilar from the statistics reported by the International Astronomical Union whose membership is 78% male[2] but the participation of female members has increased in younger age groups over time.

We often find that certain groups are excluded from the field of astronomy and astrophysics[3]. Similarly in other S20 countries, groups from rural areas, economically disadvantaged households and other marginalized groups are often left out of the astronomy and astrophysics ecosystem. In striving to uphold the principles of inclusivity and diversity, it is essential that our approach encompasses the participation and perspectives of all individuals. This commitment to inclusivity is vital for fostering a collaborative and diverse scientific community. Only by being fully inclusive and open to the diversity of thought that comes from diversity of backgrounds and lived experiences can we truly innovate and meet the challenges in this field. Reaching such an equitable outcome would certainly lead to sustainable economic growth for all[4].

The International Astronomical Union (IAU) Office of Astronomy for Development (OAD) plays a significant role globally in assisting countries in achieving objectives like the UNSDG-8. The office focuses on leveraging astronomy and its associated skills, experts, and facilities by marshalling essential human and financial resources. This is mainly done to unlock the field's scientific, technological, and cultural contributions to society. The primary method for achieving these aims is through the financial backing and organization of projects that employ astronomy as an instrument for tackling sustainable development challenges, including workforce development.

Similarly, professional astronomy and physics societies (e.g., Astronomical Society of India, American Astronomical Society, European Astronomical Society, African Astronomical Society, etc.) play an important role in bringing together professionals at all career stages and advancing the field of astronomy and astroinformatics. These organizations offer valuable expert opinion to governmental bodies and international organizations, facilitating the development of global collaborations and partnerships.

---

[1] Mulvey, P., & Pold, J. (2023, January). *New Astronomy PhDs: What Comes Next* | American Institute of Physics. https://www.aip.org/statistics/reports/new-astronomy-phds-what-comes-next-181920
[2] International Astronomical Union (no date) IAU. Available at: https://www.iau.org/administration/membership/individual/distribution/ (Accessed: 12 November 2023).
[3] For example, in South Africa, the number of black South Africans in astronomy is growing but remains small and disproportionate from the population. In the United States, the fraction of Black, Hispanic, or Native American astronomers is < 2% even though these groups make up ~30% of the US population.
[4] United Nations (UN) Sustainable Development Growth #8

**Science, Technology, Engineering, Mathematics (STEM)**
STEM education is critical to the future of astronomy and astroinformatics. It is important to encourage students to pursue STEM education by making the subject matter accessible and engaging. Students must be able to envision themselves as successful participants in the endeavour[5]. As a first step, astronomy clubs, science fairs, science museums and planetariums can play a pivotal role in broadly expanding the reach of astronomy to the public. Experience shows that such broadening participation efforts, while of great interest to the general community, are inefficient at significantly improving participation from historically excluded groups. Research shows that both the cultural context for the work, as well as representation at all levels, are equally important for attracting the best talent into a field[6]. Inclusion must be coupled with diversity to lead to better representation and subsequent innovation in a field. To accomplish this, professionals and educators need time, resources, and support to effectively bring people to STEM careers. In addition to this, special attention needs to be paid to find effective and novel solutions to systemic barriers that prevent the full participation of the best talent from all demographics into STEM.

Unlike other sciences that require hands-on experiments for data collection, astronomy has an unmatched potential for reaching a wide population because the data is often collected from remote ground- and space-based observatories and can be analysed with distributed computing resources. Moreover, open data policies and relatively standardized data formats (e.g., FITS) allows the possibility for reaching audiences in remote areas with massive on-line courses. Astronomy outreach could also be accomplished through popular science books, documentaries, educational outreach programs covering basic to higher-level years and other media[7]. While outreach and education are important first steps, well-thought out and intentional efforts after the initial outreach are important to address the systematic barriers to inclusion noted above.

**Removing Barriers to STEM Education and Career Pathways**
At each step in the education-career pipeline, it is necessary to identify critical barriers that disproportionately impact historically excluded groups. In the present context, it is crucial to approach the identification of barriers with a diverse and inclusive perspective. Historically, this task has frequently been undertaken by individuals in positions of privilege and resources. To ensure a more equitable and effective process, it is essential to involve a broad range of stakeholders, particularly those directly affected by these barriers, in both identifying and addressing them. This approach will contribute to more nuanced and effective solutions that are grounded in the realities of those most impacted. We have to turn this usual practice around and begin the task of inviting the historically excluded by listening to them and empowering them to come up with their own solutions that work in their context for increased participation in STEM.

---

[5] Permanasari, A., Rubini, B., & Nugroho, O. F. (2021). STEM education in Indonesia: Science teachers' and students' perspectives | Journal of innovation in educational and cultural research. https://doi.org/10.46843/jiecr.v2i1.24

[6] Transforming enterprises through diversity and inclusion. (2022, April 6). International Labour Organization. https://www.ilo.org/actemp/publications/WCMS_841348/lang--en/index.htm

[7] National Research Council. 2001. Astronomy and Astrophysics in the New Millennium. Washington, DC: The National Academies Press. https://doi.org/10.17226/9839.

**Increasing Inclusion & Diversity**
An inclusive environment in which a diverse workforce can thrive is critical for a successful and innovative workforce. There are several strategies that can be employed to increase inclusion and diversity in astronomy and astroinformatics:
- Specific programs that begin with an initial outreach towards historically excluded and/or underrepresented groups should be designed with the needs and barriers of those excluded in mind. To enhance participation in STEM fields, it is crucial to transition from traditional models that primarily focus on direct knowledge transfer to more inclusive approaches. These approaches should be tailored to understand and address the unique motivations and interests of individuals or groups from historically underrepresented groups. In this manner, we can foster an environment that encourages and also values diverse perspectives and contributions in STEM. It is imperative that teams foster a culture that actively cultivates a sense of belonging. Efforts such as regular assessments and, if needed, strategic adjustments to affirm an environment that is inclusive and welcoming for all members may be pursued with the same level of dedication and commitment that is customarily applied to advancing scientific research, underscoring the parallel importance of both inclusive culture and scientific progress in achieving our collective goals.
- The next step after retention is ensuring those from historically excluded groups can thrive in the field. Good mentors, champions, allies are critical in this regard. Students from historically excluded groups are often in the minority and care should be taken to avoid tokenism or feelings of being a sole representative of a demographic - thus cluster recruitment and hiring, dual mentoring models, mentor pod models can be regularly employed.
- Scholarships and Financial Support: The major obstacle for students from economically disadvantaged groups is the capital to support academic and skill education is the capital support. A commitment to paying a fair wage for all work is critical especially at entry levels of STEM. Similarly, knowledge of finances, tips on budgeting, managing resources, future financial returns in STEM fields (i.e., salaries, benefits) are all important pieces of information one must provide to early career students, especially those from impoverished backgrounds.
- Programs for Identification of Hidden Biases: It is also important to train those who are not from historically excluded groups to learn how to recognize their own bias and actively work against them. Experiences with double blind reviewing processes at NASA, NIH, and in various time allocation committees have shown increases in the diversity of successful proposers, particularly increasing the number of first-time proposers and early career researchers. However dual anonymous or double-blind reviews cannot fundamentally change the demographics of a field and that requires more active efforts such as those described above.

**Training and Professional Development**
Training and professional development opportunities are critical for building a skilled workforce in astronomy and astroinformatics. There are several strategies that can be employed to provide training and professional development opportunities to students and early-career professionals:
- Internship Programs: Paid internship programs can provide students with practical experience in the field of astronomy and astroinformatics. These programs can be designed to provide students with exposure to a wide range of skills and techniques and could possibly be done remotely for communities that are difficult to reach. It is critical to add longitudinal mentoring and follow up with such programs to ensure that interns are able to translate their experiences into the next phases of their scientific or technical careers.

- Workshops and Conferences: Workshops and conferences could be used to provide training and professional development opportunities to early-career professionals. Free or relatively inexpensive participation for those early in their careers, especially those from historically excluded groups, is an effective way to grow the broad talent in a field. At these events proactive and intentional professional development of students to ensure collaboration and networking amongst professionals in the field can be encouraged. Such conferences and networks can also proactively and intentionally focus on non-academic and non-astronomy careers to showcase the wide range of careers that are possible with a STEM degree. Meeting with astronomy graduates who work in non-astronomy field could also give wider perspective for students who consider to study astronomy at the university.
- Mentorship Programs: Mentorship is critical for early-career professionals - it can provide guidance and support from experienced professionals in the field. Mentorship programs can be designed to teach professionals the basics of good mentoring, teach mentees the elements of being a good mentee as well as promote novel mentoring models with multiple mentors and mentees in two to six person pods. Accountability for both the mentors and mentees is critical for successful programs.
- Outreach through collaboration software and citizen science is a powerful tool to increase awareness of the importance of observational astronomy and to engage young students that may decide to join a scientific program later. Successful citizen science applications could be deployed on a larger scale, ideally through school programs, to increase awareness of the challenges and possibilities of astronomical research to K-12 and equivalent levels schools. In some cases (especially when students in one country are participating in a project from another country and the students/interns cannot be paid), this may be the only way. Therefore, better guidelines for academic credits or certifications can be developed. Remote participation can be helpful for students from developing countries in providing access to a wider range of topics of study.

**Conclusion**

In order to ensure the continued success of astronomy and astroinformatics, it is critical to develop an inclusive, diverse and skilled workforce. This may be accomplished through a combination of a focus on STEM education, intentionality in and, increasing diversity at all levels of the field through innovative practices addressing systemic barriers, as well as, providing training, professional development, mentoring and championing opportunities to students and early-career professionals.

**S20 Co-Chair**: Ashutosh Sharma, Indian National Science Academy
**INSA S20 Coordination Chair:** Narinder Mehra, Indian National Science Academy

**Contributors**
Kartik Sheth, NASA, USA
Kevin Govender, Office of Astronomy for Development, South Africa
Vanessa McBride, Office of Astronomy for Development, South Africa
Laura Trouille, The Adler Planetarium, USA
Puji Irawati, National Astronomical Research Institute, Thailand
Rana Adhikari, California Institute of Technology, USA
Rafael Santos, Instituto Nacional de Pesquisas Espaciais, Brazil
Paula Coehlo, University of São Paulo, Brazil
Guiseppe Longo, Università degli Studi di Napoli Federico II, Italy
Pranav Sharma, Indian National Science Acadmey, India
Ashish Mahabal, California Institute of Technology, USA

**Appendix**

**Examples of Successful Workforce Development Strategies**

There are a number of successful workforce development strategies that have been employed in astronomy and astroinformatics. Some of them are listed below:
- REU programs:
  - LIGO SURF program: ~25 students each year have been trained since ~2005 including 2-4 Indian students each summer since ~2009 for LIGO India workforce development. Includes students at Caltech, as well as LIGO US Observatories. Administered by IUCAA on the Indian side.
- The National Science Foundation's Astronomy and Astrophysics Postdoctoral Fellowship (AAPF) program provides opportunities for early-career scientists to pursue independent research in astronomy and astrophysics.
- The Astronomy Ambassadors program, developed by the American Astronomical Society, provides training and resources for astronomers to engage with the public and promote astronomy education and outreach.
- The Space Telescope Science Institute's (STScI) Summer Internship Program provides undergraduate students with the opportunity to work on research projects related to the Hubble Space Telescope.
- SARAO scholarships and fellowships programme (https://www.sarao.ac.za/students/funding/) and general HCD strategy (https://www.sarao.ac.za/about/services/strategy-and-partnerships/human-capital-development/#)
- African Astronomical Society Seed Research Grant (https://www.africanastronomicalsociety.org/afas-2023-seed-research-grant/)
- African Network of Women in Astronomy (https://www.africanastronomicalsociety.org/afnwa/)
- https://www.advance-he.ac.uk/equality-charters/athena-swan-charter
- The International Astronomical Union Office of Astronomy for Development (OAD) and the several regional OAD offices around the world support workforce development programs and opportunities, build networks within each region, and connect audiences to resources and opportunities, etc. https://www.astro4dev.org/
- https://calbridge.smapply.org/
- https://www.akamaihawaii.org/ - there are significant economic development opportunities related to the telescopes and observatories around the world. These opportunities include providing training and professional development for the local communities, often from underrepresented groups in STEM. The Akamai Program in Hawaii is one such program, building Hawai'i's scientific and technical workforce by working directly with the local communities in the vicinity of the observatories.
- Plus programs like: Hubble fellowships, National Astronomy Consortium, TAURUS, PAARE, MUCERPI, FaST9 etc.